\documentclass[prl,twocolumn,showpacs]{revtex4}

\usepackage{graphicx}
\usepackage{dcolumn}
\usepackage{bm}

\begin{document}


\title{Modulation of Noise in Submicron GaAs/AlGaAs Hall Devices by Gating}

\author{Yongqing Li}
\email{yqli@iquest.ucsb.edu}
\author{Cong Ren}
\author{Peng Xiong}
\author{Stephan von Moln\'{a}r}
\affiliation{MARTECH and Department of Physics, Florida State
University, Tallahassee, FL 32306-4351}
\author{Yuzo Ohno}
\author{Hideo Ohno}
\affiliation{Laboratory for Electronics Intelligent Systems,
Research Institute of Electrical Communication, Tohoku University,
Sendai, Japan}

\date{\today}

\begin{abstract}

We present a systematic characterization of fluctuations in
submicron Hall devices based on GaAs/AlGaAs two-dimensional
electron gas heterostructures at temperatures between $1.5$\,K to
$60$\,K. A large variety of noise spectra, from $1/f$ to
Lorentzian, are obtained by gating the Hall devices. The noise
level can be reduced by up to several orders of magnitude with a
moderate gate voltage of $0.2$\,V, whereas the carrier density
increases less than $60\%$ in the same range. The significant
dependence of the Hall noise spectra on temperature and gate
voltage is explained in terms of the switching processes related
to impurities in n-AlGaAs.
\end{abstract}

\pacs{85.30-z, 74.40+k, 73.23-b}

\maketitle

Developing high-sensitivity noninvasive techniques for nanoscale
magnetic measurement is one of the great challenges in nanoscience
today. Such instrumentation is necessary to address important
scientific questions such as the dynamics of magnetic switching of
magnetic nanoparticles \cite{Wernsdorfer00} and/or the motility
and interactions of magnetically labeled biological molecules in
biomotors and biosensors \cite{Tamanaha01}. Extensive effort has
been made to improve the sensitivity of magnetometers for
\textit{single} nanoparticle measurements by miniaturizing the
detecting devices to micron and even nanometer scale, such as a
micrometer sized SQUID loop \cite{Wernsdorfer00}, a
micro-cantilever \cite{Harris99}, and a submicron semiconductor
Hall gradiometer \cite{Li02}. Among these techniques, Hall
magnetometers based on two dimensional gas (2DEG) III/V
semiconductor heterostructures have many advantages over other
techniques such as wide range of operational temperature and
applied field \cite{kent94}. We have recently demonstrated a
moment sensitivity of $\sim10^5$\,$\mu_B$ in a perpendicular field
of $0.3$\,T by measuring a singe Fe particle of diameter
$d\sim5$\,nm with a submicron GaAs/AlGaAs 2DEG Hall magnetometer
\cite{Li02}. Further miniaturization should increase the average
stray field of the magnetic nanoparticle and result in even higher
sensitivity, if the noise can be maintained at low level.
Unfortunately, miniaturization is also expected to increase the
noise level, especially $1/f$ noise at low frequencies. Therefore,
a detailed understanding of sources of noise is critical in device
optimization and noise reduction.

There exists a long history of noise studies in a variety of
GaAs/AlGaAs 2DEG devices.  These include \textit{resistance}
fluctuations in quantum point contacts (QPC) confined by the
split-gate technique \cite{Dekker91,Timp90,
Sakamato95,Smith96,Kurdak97} as well as macroscopic scale devices
\cite{Kurdak97,Ren93b}. Several mechanisms have been proposed to
account for the observed $1/f$ noise and Lorentzian noise,
including electron trapping \cite{Dekker91}, switching in remote
impurities \cite{Timp90,Cobden92,Kurdak97}, electron screening
effect \cite{Py96}, and DX centers \cite{Smith96,Carey96}. No
definitive conclusions have, however, been reached. Surprisingly,
there are few studies of the Hall voltage noise in
GaAs/Al$_x$Ga$_{1-x}$As 2DEGs, especially at temperatures below
$77$\,K at which these Hall devices have the best performance. To
the best of our knowledge, in this temperature range, there was
only one report on Hall voltage noise of (large)
$25\times50$\,$\mu$m$^2$ Hall bars at $4.2$\,K \cite{Kurdak97}. In
submicron 2DEG devices, various mesoscopic effects proliferate,
leading to transport properties considerably different from
macroscopic ones. Although the work on QPC samples at low
temperatures has given much insight into this problem, direct
measurement of fluctuations in submicron Hall devices may provide
us further in-depth understanding of the dynamic aspect of
transport properties.

In this letter, we report the first detailed measurement of low
frequency Hall effect noise in submicron GaAs/AlGaAs 2DEG devices
at temperatures between $1.5$\,K and $60$\,K. We observed a
pronounced dependence of both the type and the magnitude of the
noise on temperature and gating. In particular, we discovered that
a significant suppression of the noise can be achieved with
moderate gate voltage.

The GaAs/Al$_x$Ga$_{1-x}$As heterostructure used in this study was
grown on an undoped GaAs substrate and consists of a $1000$\,nm
thick undoped GaAs buffer layer, a $30$\,nm thick undoped
Al$_{0.29}$Ga$_{0.71}$As spacer layer, a $100$\,nm thick Si doped
layer with a dopant density of $10^{18}$\,cm$^{-3}$ and a $10$\,nm
thick GaAs cap layer. The electron density and Hall mobility of
this wafer were determined to be
$n\simeq1.5\times10^{11}$\,cm$^{-2}$ and
$\mu_H\simeq10^5$\,cm$^2$/V$\cdot$s at $T\leq77$\,K, respectively.
Submicron Hall bar patterns were fabricated by electron beam
lithography followed by wet etching. The sample used for noise
measurements to be presented here has three Hall crosses of
$0.7\times0.7$\,$\mu$m with an etching depth of $70$\,nm and a
$50$\,nm thick Au/Cr gate deposited on top. The carrier
concentration at zero gate voltage ($V_G=0$) determined from the
Hall measurements is $\sim1.2\times10^{11}$\,cm$^{-2}$, and varies
by less than $1\%$ for $T\leq75$\,K. Similar results on a number
of other samples from the same wafer having various geometries are
given in Ref.\ \cite{Li03}.

\begin{figure}
\includegraphics[width=7cm]{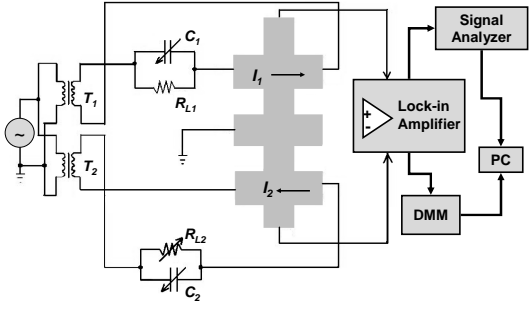}
\caption{\label{fig:HallGradi7T} The seven-terminal gradiometry
circuit for a.c.\ Hall-effect noise measurements.}
\end{figure}
A 7-terminal a.c.\ gradiometry setup shown in
Fig.~\ref{fig:HallGradi7T} was used for the noise measurement. The
two currents $I_1$ and $I_2$ applied to Hall crosses 1 and 2 are
supplied from two identical transformers, which have common input
from a function generator. Both the amplitude and the phase of
$I_1$ and $I_2$ can be balanced by limiting resistors and shunt
capacitors to ensure that the output $\Delta V_H$ of the Hall
gradiometer has a zero offset. This design was adapted from a
5-terminal ac circuit for measuring resistance fluctuations
\cite{Scofield87}. The unique feature of the present design is
that the balancing circuit is provided by passive elements, and
the symmetry ensures less vulnerability to external fluctuations.
However, the measured noise power spectral density (PSD) is
$S_{VH}=S_{VH1}+S_{VH2}$, where $S_{VH1}$ and $S_{VH2}$ are the
PSD of Hall cross 1 and 2, respectively. The output signal of a
lock-in amplifier (PAR 124A or EG$\&$G 5301) operating at
frequency $f_c\sim517$\,Hz in band-pass mode ($Q=1$) was processed
into a spectrum analyzer. We used a superconducting magnet in
persistent mode to maintain a constant field perpendicular to the
2DEG plane. All spurious sources of noise, especially those coming
from the gate itself could be ruled out or eliminated.

\begin{figure}
\includegraphics[width=6.5cm]{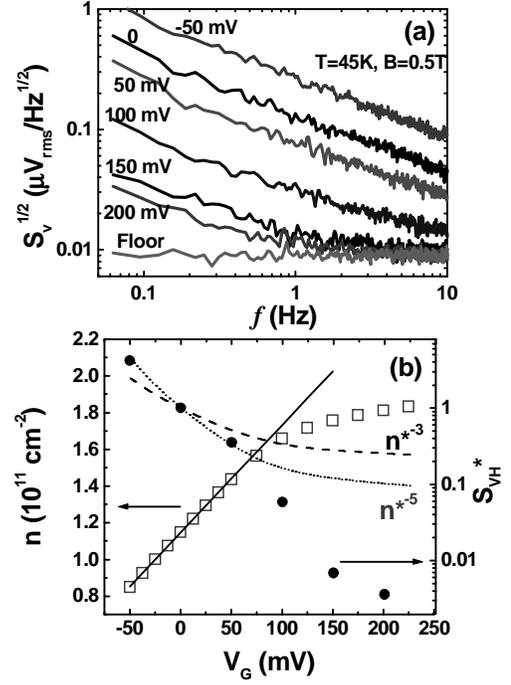}
\caption{\label{fig:Noise45K} (a) Gate voltage dependence of noise
spectra at $T=45$\,K and $B=0.5$\,T taken with $I\sim1$\,$\mu$A.;
(b) Gate voltage dependence of carrier concentration (open
squares) and the scaled noise power spectrum density
$S_{VH}^*=S_{VH}(V_G)/S_{VH}(V_G=0)$ at $1$\,Hz (solid circles).
$n^*$ is the relative change in carrier density, defined as
$n^*=n(V_G)/n(V_G=0)$.}
\end{figure}
At $V_G=0$, the noise has $1/f^\alpha$-like spectra with
$\alpha=1.0\pm0.1$ at $T=15$\,K to $60$\,K in an applied field up
to at least $B=1$\,T. In all cases, the PSD at $1$\,Hz scales with
$I^2$ as expected. Furthermore, the normalized $S_{RH}=S_{VH}/I^2$
has a linear $B^2$ dependence at all temperatures. Although the
quadratic dependence for the Hall effect noise was predicted and
observed for high mobility samples ($\mu_HB\gg1$) in the diffusive
regime \cite{Ren93a}, it is surprising that it is followed so well
in our samples where ballistic transport is not negligible.

The most dramatic observations occurred when a gate voltage was
applied to the device. Although the exact gate voltage dependence
varies at different temperatures, \textit{with} $V_G=0.2$\,V
\textit{the noise level was suppressed by more than two orders
magnitude in the entire temperature range} ($5$-$60$\,K). The
gating behavior is straightforward at high temperatures. For
example, at $T=45$\,K, $S_{VH}$ is consistently suppressed while
maintaining the $1/f$ characteristic, as shown in
Fig.~\ref{fig:Noise45K}(a). At $V_G=0.2$\,V, the noise level is
almost reduced to the background noise level at high frequencies.
The relative reduction in $S_{VH}$,
$S_{VH}^*=S_{VH}(V_G)/S_{VH}(V_G=0)$, at different gate voltages,
is plotted in Fig.~\ref{fig:Noise45K}(b), which exhibits a
decrease of a factor of $\sim300$ from $V_G=0$ to $0.2$\,V. This
reduction is much greater than what is expected from the increase
in carrier density with gating. A $n^{-3}$ dependence is expected
from Hooge's empirical rule, i.e.\
$S_{VH}=\frac{\alpha_H}{Nf^\alpha}V_H^2\propto n^{-3}$, where $N$
is the total carrier number in the device, and $\alpha_H$ is the
Hooge's constant. The measured change in carrier density $n$ of
$\sim60\%$ from $V_G=0$ to $0.2$\,V should yield only a factor $4$
decrease in $S_{VH}$, which clearly contradicts our results.
Similar gate voltage dependencies has been observed at fields from
$B=0.25$\,T to $1$\,T. The unexpected rapid decrease in $S_{VH}$
with $V_G$ is of great practical importance, since the quantity
$S_{VH}/V_H^2$ will determine the ultimate field sensitivity of
the Hall device. We point out that the noise level for a
$0.7\times0.7$\,$\mu$m$^2$ Hall cross at $T=45$\,K, $B=0.5$\,T,
and $V_G=0.2$\,V is only
$S_{RH}/2=2.5\times10^{-5}$\,$\Omega^2$/Hz at $1$\,Hz, which is
even smaller than an ungated macroscopic sample at $T=4.2$\,K
\cite{Kurdak97}.
\begin{figure}
\includegraphics[width=7.5cm]{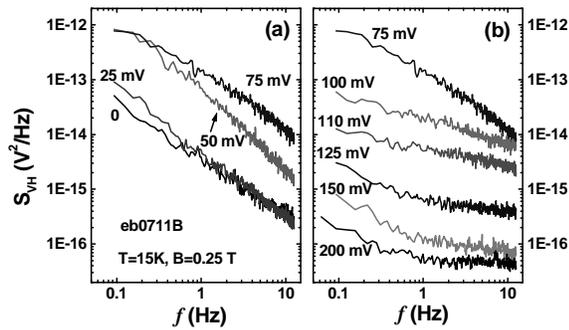}
\caption{\label{fig:PSD15K} Gate voltage dependence of noise
spectra at $T=15$\,K and $B=0.25$\,T taken with
$I\simeq1.45$\,$\mu$A. (a) Noise level increases with increasing
gate voltage at $V_G\leq75$\,mV; (b) Noise level decreases rapidly
as $V_G$ increases beyond $75$\,mV.}
\end{figure}

The gate voltage dependence of $S_{VH}$ at low temperatures is
considerably more complex than that at $T=45$\,K.
Fig.~\ref{fig:PSD15K}(a) and (b) show the results at $T=15$\,K.
Initially ($V_G\leq75$\,mV), an \textit{increase} in noise level
and deviation from $1/f$ spectra are observed with increasing
$V_G$. Further increase in $V_G$ results in a rapid reduction in
noise level and restoration of $1/f$ spectra. The complex behavior
at low temperature cannot be explained simply by changes in
carrier density or mobility since both quantities remain
relatively constant below $60$\,K. At $V_G=75$\,mV, the noise
reaches the highest level and the spectra are more Lorentzian like
than $1/f$, concomitant with the approach to saturation of $n$ at
$V_G>0.1$\,V (Fig.\,2(b)). Such a transition from linear
dependence to saturation for $n$ has been observed and explained
by Hirakawa \textit{et al.}\cite{Hirakawa84} who noted that when
$V_G$ is smaller than a certain threshold voltage ($\sim75$\,mV in
this case), the quasi-Fermi level in n-doped AlGaAs is lower than
the donor level, the Si donors are fully depleted and the gated
2DEG structure acts like a parallel plate capacitor, resulting in
the linear dependence of carrier concentration on $V_G$. However,
for larger values of $V_G$ the quasi-Fermi level passes through
the donor level, and a neutral region forms in the n-AlGaAs layer,
which leads to the saturation effect. The coincidence of the
highest noise level and the appearance of Lorentzian behavior with
the beginning of the formation of the neutral layer strongly
suggests that the switching processes related to donors in
n-AlGaAs may be responsible for the pronounced and complex gate
voltage dependence of fluctuations. Furthermore, the shielding
effect by the neutral layer at higher $V_G$ offers a natural
explanation for the substantial noise reduction if the
fluctuations are induced by trapping and emission processes in
n-AlGaAs. The key question is: how does one explain the
dramatically different gating behavior at different temperatures?

\begin{figure}
\includegraphics[width=7cm]{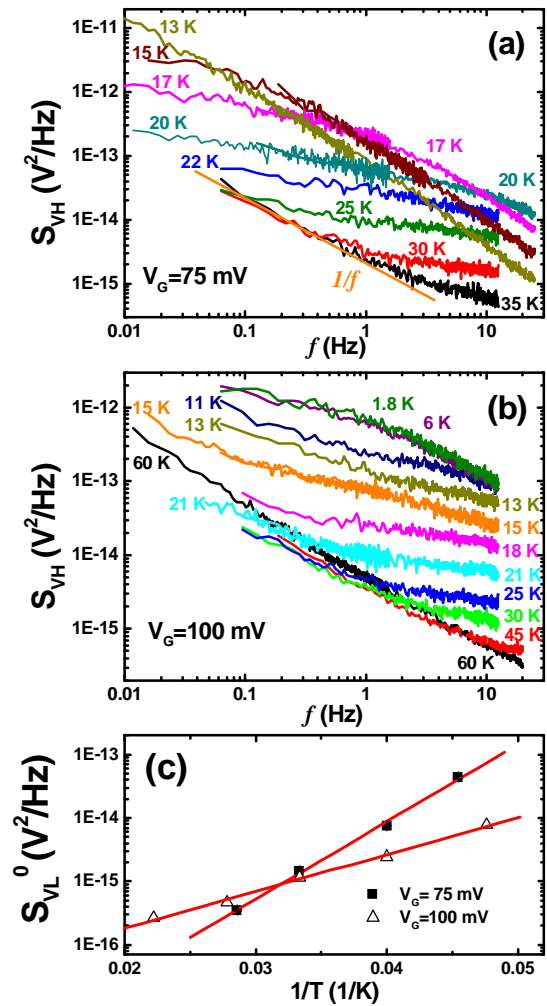}
\caption{\label{fig:Tdep} Temperature dependence of noise spectra
at (a) $V_G=75$\,mV and (b) $V_G=100$\,mV; (c) shows the
$\log(S_{VL}^0)-1/T$ fits for these two gate voltages.}
\end{figure}

Fig.~\ref{fig:Tdep}(a) shows the noise spectra taken at
$V_G=75$\,mV and $T=13$\,K to $35$\,K. With decreases in
temperature, the noise level increases rapidly and the deviation
from $1/f$ becomes clear. At higher temperatures, the noise
spectra are $1/f$-like at lower frequencies, but become flat at
higher frequencies. At $T=20$\,K, the local spectral slope,
defined as $\nu(f)\equiv-\partial\ln S_VH(f)/\partial\ln f$,
remains a value less than $1$ over the entire measured frequency
range ($0.01$-$25$\,Hz). For the spectra at $T=13$-$17$\,K,
$\nu(f)$ varies monotonically from less than $1$ to larger than
$1$. We can therefore, use the characteristic frequencies $f_p$,
at which $\nu(f)$=1, to extract the peak activation energy $E_p$.
Fitting the data to an Arrhenius law
$f_p=f_0\exp(-\frac{E_p}{k_BT})$, we obtain $E_p=27\pm3$\,meV and
attempt frequency $f_0=10^8$-$10^9$\,Hz, which is about the same
as that obtained from the telegraph noise measurements on the
GaAs/AlGaAs 2DEG QPC samples ($f_0\sim$10$^9$\,Hz)
\cite{Dekker91}. The Gaussian spreading width of the activation
energy distribution around $E_p$ obtained from $T=13$-$20$\,K data
is about $8$\,meV, based on the Dutta-Horn approach
\cite{Dutta79}. This narrow distribution in $E_a$ gives rise to
Lorentzian type of spectra
($S_{VL}(f)=S_{VL}^0/(1+4\pi^2f^2\tau^2)$ with
$S_{VL}^0\propto\tau=\tau_0\exp(E_a/k_BT)$ and $\tau_0\sim
f_0^{-1}$) which dominate at lower temperatures. At higher
temperatures, the corner frequency $f_c$($\sim\tau^{-1}$) becomes
much larger than $25$\,Hz, so the observed noise spectra can be
approximately decomposed as:
$S_{VH}(f)=\frac{S_{VH}^0}{f}+S_{VL}^0$, where the first term is
the $1/f$ component corresponding to a uniform distribution in
activation energy $E_a$, and $S_{VL}^0\propto\tau$ is the flat
part ($f\ll f_c$) of the Lorentzian spectrum. The strong
temperature dependence of $\tau$ can explain the rapid increase in
the noise level in a narrow temperature range ($35$\,K to
$20$\,K). As shown in Fig.\,\ref{fig:Tdep}(c), the
$\log(S_{VL}^0)$-$1/T$ fit for data at $T=22$-$35$\,K yields
$E_a=24$\,meV, which is reasonably close to the $E_p\simeq27$\,meV
obtained from the $T=13$-$17$\,K data. The overall feature of the
noise spectra at $V_G=100$\,mV, shown in Fig.~\ref{fig:Tdep}(b),
is similar to that at $V_G=75$\,mV, but the noise level is lower.
With a similar $\log(S_{VL}^0)$-$1/T$ fit at temperatures between
$21$\,K and $45$\,K, we obtain $E_a\simeq12$\,meV
(Fig.\,\ref{fig:Tdep}(c)). At temperatures below $6$\,K, the noise
spectra stop changing with temperature, which can be explained by
a crossover from thermal activation to quantum tunneling
\cite{Dekker91}.

So far, we have seen that the large temperature dependence of the
noise spectra at intermediate gate voltages can be attributed to
the thermally activated process related to a narrow distribution
in the activation energy $E_a$. The gate voltage dependence of the
carrier density indicates that the ionized impurities in n-AlGaAs
start to capture electrons noticeably at $V_G\sim75$\,mV and form
a neutral layer at higher $V_G$. However, at $V_G\leq0.1$\,V, the
occupied impurity density is so low that the n-AlGaAs is still
insulating and its transport is via electron hopping between
localized states. The energy barrier for the thermally activated
hopping is dependent on the overlapping of the corresponding
wavefunctions. This can explain why the extracted activation
energy from the noise spectra decreases as the gate voltage
increases. The latter increases the neutral impurity density, and
hence the overlap of the localized electron wavefunctions also
becomes larger. The large gate voltage dependence of the noise
spectra, therefore, can be attributed to the change of the thermal
activation energy as well as the neutral impurity density with the
gate voltage.

In contrast to the noise spectra at intermediate gate voltages,
the noise at $V_G=0$ and $0.2$\,V has a $1/f$-like spectrum over
the entire temperature range ($15$-$60$\,K). The frequency
exponent for each spectrum was found to be close to $1$
($0.9<\alpha<1.1$). The noise level increases with temperature,
which is expected for the thermally activated switching process
\cite{Dutta79}. At $V_G=0.2$\,V and $T=15$-$60$\,K, $S_{VH}$
maintains scaling with $I^2$ up to very large currents
($I\simeq6$\,$\mu$A), which is surprising because considerable
electron heating effect (tens of Kelvins) has been observed from
transport measurements, such as longitudinal magnetoresistance and
small non-linear component of the Hall voltages. The lattice
heating determined from noise measurements at $T=15$\,K and
$V_G=0.1$\,V is less than $0.5$\,K/$\mu$A. This indicates the
noise is almost independent of the electron temperature in our
experimental range, which rules out the possibility that the
dominating noise observed in this work comes from any effect
related to the 2DEG alone, such as the electron-electron
interaction. On the other hand, the strong dependence of the noise
on the lattice temperature further supports the picture that the
observed noise originates from the remote impurities in n-AlGaAs.

In summary, we have shown that submicron Hall devices combined
with gating are excellent probes to study the fluctuations in
GaAs/AlGaAs 2DEG heterostructures. The detailed study of the noise
has shown that the large gate voltage dependence cannot be
explained by changes in the electron density or mobility. The data
suggest that the thermally activated switching processes related
to remote impurities are responsible for the observed noise
behavior except at very low temperatures where a crossover to
quantum tunneling occurs. Most importantly, we have found that
suppression of noise up to a few orders of magnitude can be
achieved by moderate gating over the entire temperature range
($5$-$60$\,K), which is of great importance for device
applications. In fact, the gating effect has been utilized for the
measurements of individual magnetic nanoparticles with moment
sensitivity better than $\sim10^4$\,$\mu_B/\sqrt{Hz}$ at $1$\,Hz
and $T=15$\,K and a perpendicular applied field $B=0.25$\,T
\cite{Li03}. Clearly, as the dimensions of magnetic structures are
further reduced and the surface to volume ratio increases, these
techniques will afford examination of new magnetic effects due to
surface interactions.

We gratefully acknowledge stimulating discussions with B.\ Raquet,
P.\ Schlottmann, S.J.\ Bending, P.\ Stiles and R.\ S.\ Popovic.
This work was supported by NSF grant DMR0072395, and by DARPA
through ONR grants N-00014-99-1-1094 and MDA-972-02-1-0002. The
work at Tohoku University was supported partially by a
Grant-in-Aid from the Ministry of Education, Japan, and by the
Japan Society for the Promotion of Science.

\bibliography{Hall_noise}

\end{document}